\documentclass[letterpaper]{article} %
\usepackage[]{aaai2026}  %
\usepackage{times}  %
\usepackage{helvet}  %
\usepackage{courier}  %
\usepackage[hyphens]{url}  %
\usepackage{graphicx} %
\usepackage{natbib}  %
\usepackage{caption} %
\usepackage{algorithm}
\usepackage{algorithmic}
\usepackage{xcolor}

\usepackage{newfloat}
\usepackage{listings}
\DeclareCaptionStyle{ruled}{labelfont=normalfont,labelsep=colon,strut=off} %
\floatstyle{ruled}
\newfloat{listing}{tb}{lst}{}
\floatname{listing}{Listing}
\definecolor{bgcolor}{RGB}{240,240,240}

\usepackage{fontawesome5}

\usepackage{multirow}
\usepackage{graphicx}
\usepackage{subcaption}
\usepackage{listings}
\usepackage{stmaryrd}
\usepackage{graphicx}

\usepackage{listings}
\usepackage{amsmath}

\newcommand{\gymenv}{QueryGym}
\newcommand{\nltoquery}{NL2Query}

\newcommand{\hide}[1]{}

\newcommand{\nexplore}{12}
\newcommand{\nperform}{8}

\newcommand\YAMLcolonstyle{\color{red}\mdseries}
\newcommand\YAMLkeystyle{\color{black}\bfseries}
\newcommand\YAMLvaluestyle{\color{blue}\mdseries}

\makeatletter

\newcommand\language@yaml{yaml}

\expandafter\expandafter\expandafter\lstdefinelanguage
\expandafter{\language@yaml}
{
  keywords={true,false,null,y,n},
  keywordstyle=\color{darkgray}\bfseries,
  basicstyle=\YAMLkeystyle,                                 %
  sensitive=false,
  commentstyle=\color{purple}\ttfamily,
  stringstyle=\YAMLvaluestyle\ttfamily,
  moredelim=[l][\color{orange}]{\&},
  moredelim=[l][\color{magenta}]{*},
  moredelim=**[il][\YAMLcolonstyle{:}\YAMLvaluestyle]{:},   %
  morestring=[b]',
  morestring=[b]",
  literate =    {---}{{\ProcessThreeDashes}}3
                {>}{{\textcolor{red}\textgreater}}1     
                {|}{{\textcolor{red}\textbar}}1 
                {\ -\ }{{\mdseries\ -\ }}3,
}

\lst@AddToHook{EveryLine}{\ifx\lst@language\language@yaml\YAMLkeystyle\fi}
\makeatother

\newcommand\ProcessThreeDashes{\llap{\color{cyan}\mdseries-{-}-}}

\newenvironment{longlisting}{\captionsetup{type=listing}}{}
\usepackage[cachedir=.]{minted}

\makeatletter
\renewcommand{\paragraph}{%
  \@startsection{paragraph}{4}%
  {\z@}{0.5ex \@plus 1ex \@minus .1ex}{-2em}%
  {\normalfont\normalsize\bfseries}%
}
\makeatother

\title{\raisebox{-0.8em}{\includegraphics[width=0.05\textwidth]{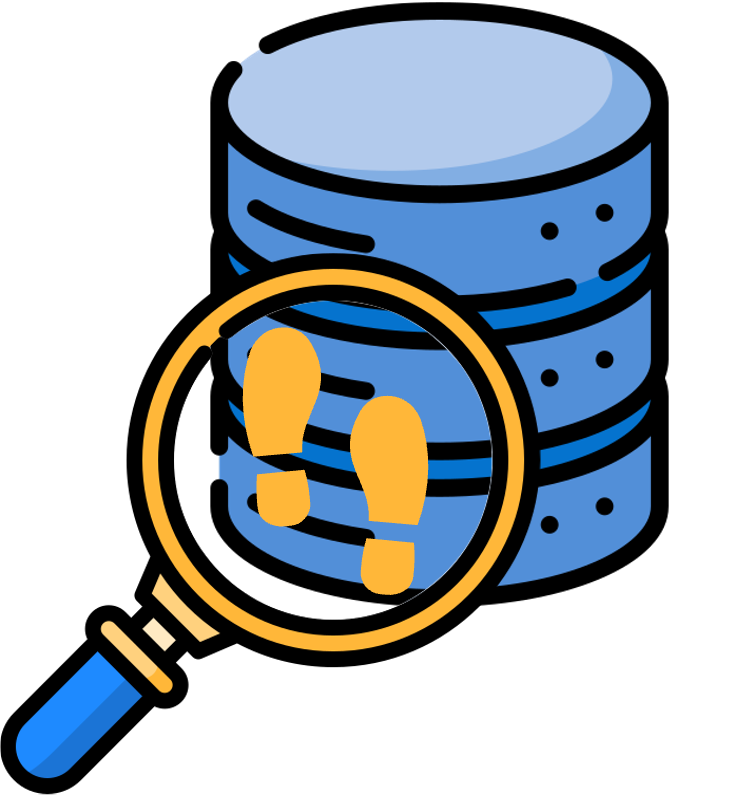}}  QueryGym: Step-by-Step Interaction with Relational Databases}
\author {
Haritha Ananthakrishnan\textsuperscript{\rm 1}, 
Harsha Kokel\textsuperscript{\rm 1},
Kelsey Sikes\textsuperscript{\rm 2}, \\
Debarun Bhattacharjya\textsuperscript{\rm 1},
Michael Katz\textsuperscript{\rm 1}, 
Shirin Sohrabi\textsuperscript{\rm 1},
Kavitha Srinivas\textsuperscript{\rm 1}
}
\affiliations{
    \textsuperscript{\rm 1}IBM Research\\
    \textsuperscript{\rm 2}Colorado State University\\
}

\usepackage{bibentry}

\begin{document}

\maketitle

\begin{abstract}
We introduce \textbf{QueryGym}, an interactive environment for %
{building, testing,}
and evaluating LLM-based query planning agents. 
{Existing frameworks often tie agents to specific query language dialects or obscure their reasoning; QueryGym instead requires agents to construct explicit sequences of relational algebra operations, ensuring engine-agnostic evaluation and transparent step-by-step planning.}
The environment is implemented as a Gymnasium interface that supplies observations---including schema details, intermediate results, and execution feedback---and receives actions that represent database exploration (e.g., previewing tables, sampling column values, retrieving unique values) as well as relational algebra operations (e.g., filter, project, join). %
We detail the motivation and the design of the environment.
{In the demo, we showcase the utility of the environment by contrasting it with contemporary LLMs that query databases}. 
{
QueryGym serves as a practical testbed for research in error remediation, transparency, and reinforcement learning for query generation. %
}
\end{abstract}

\begin{links}
    \link{Video}{https://ibm.biz/QueryGym}
\end{links}

\section{Introduction}

The ability to translate natural language questions into executable database queries---%
{across SQL and other query languages}---has become a standard benchmark for evaluating an LLM's ability to reason over relational databases. Recent advances in LLMs' reasoning capabilities have dramatically improved performance of \nltoquery{} benchmarks such as Spider and BIRDBench~\cite{spider20,LiHQYLLWQGHZ0LC23,granado2025raisereasoningagentinteractive,wang2025macsqlmultiagentcollaborativeframework,ChaseSQL,papicchio2025think2sqlreinforcellmreasoning}. However, a systematic understanding of how these models plan, explore, and manipulate relational data remains limited. Existing evaluation suites typically treat query generation as a static sequence‑to‑sequence problem, providing a single gold‑standard SQL string and scoring the final output.  
{This paradigm obscures intermediate reasoning, blocks error analysis, and rules out reinforcement learning methods that depend on an interactive agent-database loop}.

\begin{figure}
    \centering
    \includegraphics[width=\columnwidth]{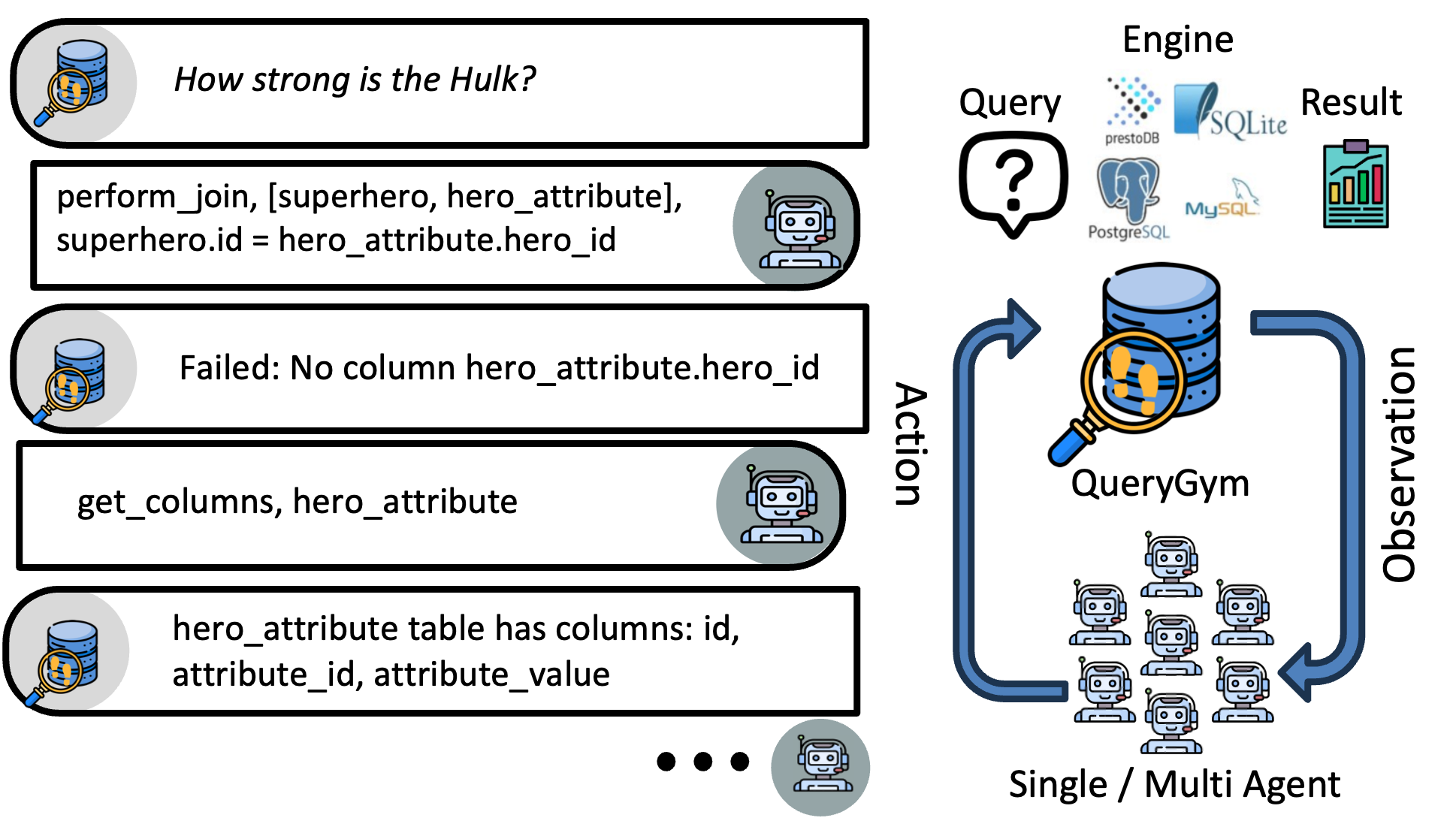}
    \caption{Sample trajectory in QueryGym, an interactive environment for query planning agents. }
    \label{fig:querygym}
\vspace*{-0.2cm}
\end{figure}

To address these gaps, we introduce \textbf{\gymenv{}}, an interactive, engine-agnostic environment that casts %
\nltoquery{} tasks
as a partially observable Markov Decision Process (POMDP).  In \gymenv{}, an agent must interact with the database
to answer a natural language question; see Fig.~\ref{fig:querygym} for an example. %
The environment supplies rich textual observations—including schema metadata, intermediate table previews, and error feedback---while accepting actions that correspond to both exploratory probes (e.g., previewing a table, sampling column values) and relational algebra manipulations. By abstracting away any concrete query language (e.g., SQL, PostgreSQL), \gymenv{} enables truly engine‑agnostic research, %
and its interactive database access naturally supports use‑cases such as error remediation.
Together, these features make QueryGym a versatile testbed for exploring and improving LLM-based query planning agents in a controlled, interactive setting.

\section{\gymenv{} Environment}

We cast %
\nltoquery{} tasks (as instantiated in SQL or other query languages)
as a POMDP tuple $(S, A, \Omega, T, R)$, where $S$ is the set of environment states, $A$ is the set of actions, $\Omega$ is the set of observations, $T: S \times A \rightarrow S $ is the deterministic transition function between states, and $R: S\times A \rightarrow \mathcal{R}$ is the reward function.

\paragraph{Environment States} The state in \gymenv{} comprises of the full schema 
and database contents, %
the natural language question, %
and any intermediate tables, Common Table Expressions (CTEs) that the agent has materialized so far.  

\paragraph{Actions} The action space consists of \nexplore{} exploration operations (such as \texttt{preview\_table} and  \texttt{get\_sample\_values}) that probe the database, and \nperform{} relational algebra operations (such as \texttt{perform\_filter}, \texttt{perform\_join}, and \texttt{perform\_union}) that manipulate the tables. Each of these operations is associated with a fixed number of parameters. The full catalog of the available operations and their required parameters is listed in the Appendix. At each turn, the agent is expected to issue a command that %
defines the chosen operation to be performed and %
supplies concrete values for all its parameters.

\paragraph{Observation} %
The textual observation supplied by the \gymenv{}  at any given turn depends on the environment state and the previous command. They can be broadly categorized into $4$ classes: \emph{Overview}, the information about the current database schema and the natural language question; \emph{Exploration Result}, the information requested by the explore operation;
\emph{Intermediate CTE Info}, 
the intermediate table that was created as a result of the last relational algebra operation; \emph{Error Feedback}, error trace of any failure that occurred when attempting to apply a relational algebra operation, such as a missing column or unknown keyword. 

\paragraph{Transition Function} Currently, the transition function in \gymenv{} treats the underlying database 
as immutable. The exploration actions do not modify the state of the environment; they only furnish the requested information in the observation as text. For relational algebra operations, the transition function constructs a new CTE if the operation is successful, otherwise the error is passed to the observation and the state of the environment remains the same.

\paragraph{Reward Function} An episode terminates as soon as the current CTE is \emph{equivalent} to the target table, i.e., it contains exactly the same set of rows and columns, irrespective of ordering and column names. A large terminal reward is provided upon termination, whereas a small reward is provided when the CTE is a subset or superset of the target table.

\paragraph{Scope}
\gymenv{} currently supports translation of the relational algebra operations to SQLite and PostgreSQL but other SQL dialects could be easily supported by implementing a single class compatible with Gym API. Our system also automatically converts any \nltoquery{} dataset into a POMDP, %
if provided in the BIRDBench format. %
Benchmarks supported by our system include: BIRDBench~\cite{LiHQYLLWQGHZ0LC23}, WikiSQL~\cite{WikiSQL}, Criteria2SQL~\cite{Criteria2sql}, ACL SQL~\cite{KaoshikPRAJ021}, CoSQL~\cite{CoSQL}, FIBEN~\cite{FIBEN},  SEOSS-Queries~\cite{TOMOVA2022108211}, SParC~\cite{SParC}, Spider-Syn~\cite{SpiderSyn}, SQUALL~\cite{SQUALL} and BIRD-Critic (birdsql/bird-critic-1.0-postgresql)~\cite{birdcritic}.

\paragraph{Agent}
We also provide a sample implementation of a LangChain‑based Gymnasium agent that leverages a vLLM‑hosted LLM to solve the task by iteratively generating actions from textual observations, handling error feedback, and progressively constructing the correct relational algebra plan until the final CTE matches the target table.

\section{Use Cases}
\gymenv{} 
supports the following research use cases. 

\paragraph{Engine Agnostic Research} \gymenv{} enables researchers to develop, train, and evaluate query planning agents without being tied to any particular SQL dialect or database system. This design supports cross‑engine comparisons, transfer‑learning studies, and the creation of truly portable query‑generation strategies that can be deployed on SQLite, PostgreSQL, MySQL, or any future relational engine with minimal adaptation.

\paragraph{Database Exploration} Current \nltoquery{} pipelines rely on schema linking, where relevant tables and columns are first identified given the question before generating a query. Instead of depending on this schema linking step, \gymenv{} provides inexpensive probing actions that allow an agent to engage in active data exploration, akin to human analysts. These exploratory operations serve two complementary purposes. First, an agent can use data exploration as an alternative to schema linking. 
Second, an agent can disambiguate ambiguous names. For example, a column named \textit{date} in \textit{employee} table could refer to either \textit{birthdate} or \textit{joining date}; by examining the column's data distribution, an agent can resolve such ambiguity. Exploratory actions foster transparent and interpretable agents in general, enabling them to justify why a particular query plan was chosen.

\paragraph{Error Remediation}  In many real-world scenarios, users may provide an initial SQL query that requires syntactic or semantic corrections. Recent benchmarks such as BIRD-Critic~\cite{birdcritic} have emerged to evaluate LLMs’ ability to perform such debugging tasks. In \gymenv{}, the initial SQL query is treated as the first action, and the environment generates CTEs and/or feedback observations. The agent then iteratively refines the query, progressively reducing the gap between the current CTE and the target answer table until the episode concludes successfully.

\paragraph{Reinforcement Learning} The verifiable reward signal implemented in the \gymenv{} can be plugged directly into \textit{RL with Verifiable Results} (RLVR)
pipelines and an agent can be trained to reason and explore before answering the query. Moreover, the exploration trajectories can also be used for offline RL or imitation learning.

\section{Demonstration Plan}
The QueryGym demonstration consists of three components: an exploration interface, a blackbox SQL execution and analysis interface, and an agent interface. The exploration interface allows users to select example queries, inspect database schemas, and interactively execute actions such as retrieving tables or performing joins and filters. The blackbox SQL execution interface highlights the drawbacks of single-pass \nltoquery{} generation by showing how a query produced by a large LLM can fail at a single step, rendering the entire generation useless, whereas breaking it down into actions enables targeted remediation. The agent interface then introduces an agent that generates stepwise plans using the environment actions and observations, presenting trajectories both as chat style replays and flow diagrams. In our demo, we use Llama 3.3-70b Instruct \cite{grattafiori2024llama3herdmodels} as the LLM of choice.

\bibliography{custom}

\appendix

 \onecolumn 

\section{Actions}

The environment allows $\nexplore$ exploration actions and $\nperform$ relational algebra operations. The complete list of actions is presented in Listing~\ref{lst:actions}.

\begin{longlisting}
\begin{minted}[frame=single,   framesep=3mm,  xleftmargin=1pt,breaklines,   tabsize=4, fontsize=\scriptsize]{js}
  GET_OVERVIEW:
    description: Get an overview of the task.
    usage: >
        Action: ["get_overview"]
        
  GET_QUERY:
    description: Get the user query.
    usage: >
        Action: ["get_query"]

  GET_SCHEMA:
    description: Get the schema.
    usage: >
        Action: ["get_schema"]

  GET_TABLES:
    description: Get a list of tables in the database.
    usage: >
        Action: ["get_tables"]

  GET_COLUMNS:
    description: Get a list of columns in a given table.
    usage: >
        Action: ["get_columns", "<table_id>"]
    examples_description: >
      Example #1: ["get_columns", "T_0"]
      Example #2: ["get_columns", "frpm"]

  GET_ACTIONS:
    description: Get a list of actions.
    usage: >
        Action: ["get_actions"]

  GET_OPERATIONS:
    description: Get a list of operations.
    usage: >
        Action: ["get_operations"]

  PREVIEW_TABLE:
    description: Preview the first 5 rows of a table.
    usage: >
        Action: ["preview_table", "<table_id>"]
    examples_description: >
      Example #1: ["preview_table", "`employee name` as emp"]
      Example #2: ["preview_table", "crime"]

  GET_COLUMN_STATS:
    description: Get descriptive statistics of a column, it include the central tendency, dispersion and shape of a dataset’s distribution, excluding NaN values.
    usage: >
        Action: ["get_column_stats", "<table_id>", "<column_id>"]
    examples_description: >
      Example #1: ["get_column_stats", "`employee name` as emp",  "emp.address"]
      Example #2: ["get_column_stats", "crime", "crime.case_number"]

  GET_UNIQUE_VALUES:
    description: Get unique values from a specified column in a table.
    usage: >
        Action: ["get_unique_values", "<table_id>", "<column_id>"]
    examples_description: >
      Example #1: ["get_unique_values", "`employee name` as emp",  "emp.address"]
      Example #2: ["get_unique_values", "crime", "crime.case_number"]

  GET_COLUMN_TYPES: 
    description: Preview the data types of all columns in a table.
    usage: >
        Action: ["get_column_types", "<table_id>"]
    examples_description: >
      Example #1: ["get_column_types", "`employee name` as emp"]
      Example #2: ["get_column_types", "crime"]

  GET_SAMPLE_VALUES: 
    description: Get sample values from a specific column. 
    usage: >
        Action: ["get_sample_values", "<table_id>", "<column_id>"]
    examples_description: >
      Example #1: ["get_sample_values", "employee", "employee.phone_number"]
      Example #2: ["get_sample_values", "Solution as T_1", "T_1.SolutionID"]

  PERFORM_PROJECTION:
    description: Project columns from a table. 
    usage: >
        Action: ["perform_projection", "<table_id>", "<columns>"]
    examples_description: >
      Example #1: ["perform_projection", "T_0", "DISTINCT T_0.movie_title, T_0.movie_popularity"]
      Example #2: ["perform_projection", 'T_3', "AVG(T_3.rating_score) as score"]
      Example #3: ["perform_projection", "employee", "employee.`home address`"]
      
  PERFORM_FILTER:
    description: Filter rows from a table based on some condition. 
    usage: >
        Action: ["perform_filter", "<table_id>", "<conditions>", "[projected columns]"]
    examples_description: >
        Example #1: ["perform_filter", "T_1", "T_1.list_creation_date_utc >= 2016-02-01 AND T_1.list_creation_date_utc <= 2016-02-29 AND T_1.user_eligible_for_trial = 1"]
        Example #2: ["perform_filter", "T_2", "T_2.movie_release_year = 2021 AND T_2.director_name = 'Steven Spielberg'"]

  PERFORM_JOIN:
    description: >
      This action joins two tables based on a condition. When executing this action, remember to follow the exact format provided in usage.
    usage: >
      Action: ["perform_join", "<list-of-tables>", "<list-of-join-conditions>", "<list-of-join-type>", "<projected columns>"]
    examples_description: >
      If you have N tables, you should have a list of N - 1 join conditions.
      Example #1: ["perform_join", ["Table_0", "Table_1","Table_2"], ["Table_0.column = Table_1.column AND Table_1.`planet name` == 'earth'", "Table_1.`planet name` = Table_2.`planet name` and Table_2.solar_system = `Milky Way`"], ["INNER JOIN", "LEFT JOIN"], '(Table_1.`planet name` || " " || Table_2.solar_system) AS planet_location']
      Example #2: ["perform_join", ["job as T2", "job_details as T1"], ["T1.info = T2.info"], ["INNER JOIN"], "T2.OfficeCoordinates"]


  PERFORM_ORDER_BY:
    description: Order table entities based on a condition.
    usage: >
        Action: ["perform_order_by", "<table-id>", "<order-condition>", "[projected columns]"]
    examples_description: >
        Example #1: ["perform_order_by", "T_0", "(CAST(`Free Meal Count (K-12)` AS REAL) / `Enrollment (K-12)`) DESC"] 
        Example #2: ["perform_order_by", "movies", "movie_popularity DESC NULLS LAST"]

  PERFORM_LIMIT:
    description: Limit the table rows based on a condition.
    usage: >
        Action: ["perform_limit", "<table-id>", "<integer>", "[projected columns]"]
    examples_description: >
        Example #1: ["perform_limit", "`employee name` as emp", "5"]
        Example #2: ["perform_limit", "T_0", "3"]


  PERFORM_AGGREGATE:
    description: Used when results are to be aggregated or grouped by some column.
    usage: >
        Action: ["perform_aggregate", "<table-id>", "<column-id>", "<projected columns>", "[having-condition]"]
    examples_description: >
        Example #1: ["perform_aggregate", "movies", "director_name,  director_country", "director_name,  director_country", "COUNT(movies) > 3"]
        Example #2: ["perform_aggregate", "employee", "employee_id", "employee_id", "MAX(SUBSTR(list_creation_date_utc, 1, 4)) - MIN(SUBSTR(list_creation_date_utc, 1, 4)) >= 10"]

  
  PERFORM_UNION:
    description: Used to union two tables.
    usage: >
        Action: ["perform_union", "ALL|DISTINCT",  "<table-1-id>", "<table-2-id>", "[table-1-columns]", "[table-2-columns]"]
    examples_description: >
        Example #1: ["perform_union", "ALL",  "movies", "ratings", "movie.rating, movie.name", "ratings.movie_score, ratings.movie_title"]
        Example #2: ["perform_union", "DISTINCT",  "T_0", "T_2", "T_0.name", "T_2.reviews"]

  PERFORM_INTERSECT:
    description: Used to intersect two tables.
    usage: >
        Action: ["perform_intersect", "<table-1-id>", "<table-2-id>", "[table-1-columns]", "[table-2-columns]"]
    examples_description: >
        Example #1: ["perform_intersect", "movies", "ratings", "movie.rating, movie.name", "ratings.movie_score, ratings.movie_title"]
        Example #2: ["perform_intersect", "T_0", "T_2", "T_0.name", "T_2.reviews"]
\end{minted}
\caption{Complete list of actions in the gym environment, their descriptions, usage information as well as example of use.}
\label{lst:actions}
\end{longlisting}

\end{document}